\documentclass[prl,twocolumn,preprintnumbers,amsmath,amssymb,superscriptaddress,longbibliography]{revtex4-1}

\usepackage{graphicx}
\usepackage{dcolumn}
\usepackage{bm}
\usepackage{float}
\usepackage{hyphenat}
\usepackage[dvipsnames]{xcolor}
\usepackage{physics}
\usepackage[T1]{fontenc}
\usepackage{lineno}

\newcommand{\heriotwatt}{Institute of Photonics and Quantum Sciences, SUPA, Heriot-Watt University, Edinburgh EH14 4AS, United Kingdom}
\newcommand{\TsukubaKenji}{Research Center for Functional Materials, National Institute for Materials Science, 1-1 Namiki, Tsukuba 305-0044, Japan}
\newcommand{\TsukubaTakashi}{International Center for Materials Nanoarchitectonics, National Institute for Materials Science,  1-1 Namiki, Tsukuba 305-0044, Japan}
\newcommand{\Imperial}{Departments of Materials and Physics and the Thomas Young Centre for Theory and Simulation of Materials, Imperial College London, South Kensington Campus, London SW7 2AZ, United Kingdom}

\begin{document}

\title{Exciton-polarons in the presence of strongly correlated electronic states in a MoSe$_2$/WSe$_2$ moir{\'e} superlattice}

\author{Aidan J. Campbell}
\affiliation{\heriotwatt}
\author{Mauro Brotons-Gisbert}
\affiliation{\heriotwatt}
\author{Hyeonjun Baek}
\affiliation{\heriotwatt}
\author{Valerio Vitale}
\affiliation{\Imperial}
\author{Takashi Taniguchi}
\affiliation{\TsukubaTakashi}
\author{Kenji Watanabe}
\affiliation{\TsukubaKenji}
\author{Johannes Lischner}
\affiliation{\Imperial}
\author{Brian D. Gerardot}
\email{B.D.Gerardot@hw.ac.uk}
\affiliation{\heriotwatt}
    
\date{\today}

\begin{abstract}
Two-dimensional moir{\'e} materials provide a highly tunable platform to investigate strongly correlated electronic states. Such emergent many-body phenomena can be optically probed in moir{\'e} systems created by stacking two layers of transition metal dichalcogenide semiconductors: optically injected excitons can interact with itinerant carriers occupying narrow moir{\'e} bands to form exciton-polarons sensitive to strong correlations. Here, we investigate the behavior of excitons dressed by a Fermi sea localised by the moir{\'e} superlattice of a molybdenum diselenide (MoSe$¬_2$) / tungsten diselenide (WSe$¬_2$) twisted hetero-bilayer. At a multitude of fractional fillings of the moir{\'e} lattice, we observe ordering of both electrons and holes into stable correlated electronic states.  Magneto-optical measurements reveal extraordinary Zeeman splittings of the exciton-polarons due to exchange interactions in the correlated hole phases, with a maximum close to the correlated state at one hole per site. The temperature dependence of the Zeeman splitting reveals antiferromagnetic ordering of the correlated holes across a wide range of fractional fillings. Our results illustrate the nature of exciton-polarons in the presence of strongly correlated electronic states and reveal the rich potential of the MoSe$¬_2$/WSe$¬_2$ platform for investigations of Fermi-Hubbard and Bose-Hubbard physics.

\end{abstract}

\maketitle

\section{Introduction}

Two-dimensional (2D) materials have emerged as a new playground to investigate many-body interactions and strongly correlated electronic phenomena. For example, due to a direct bandgap \cite{splendiani2010emerging}, huge exciton binding energies \cite{chernikov2014exciton}, and straightforward control of carrier concentration \cite{baugher2014optoelectronic}, monolayer transition metal dichalcogenides (TMDs) provide a platform to probe the interaction of an exciton with a Fermi sea (2D electron or hole gas) described by the Fermi-polaron model \cite{sidler2017fermi}. With increasing Fermi energy, a neutral exciton evolves into two branches due to both attractive (lower energy) and repulsive (higher energy) interactions with charge carriers \cite{suris2001excitons,back2017giant,sidler2017fermi,efimkin2017many,glazov2020optical}. By extension, TMD moir{\'e} heterostructures provide access to a highly tunable many-body physical system consisting of an exciton dressed by a Fermi sea which forms a series of charge-ordered (Mott insulating and generalised Wigner crystal) electronic states as the carrier concentration is tuned \cite{shimazaki2020strongly}. 

Stacking two monolayer TMDs with either a lattice mismatch and/or relative twist angle forms a moir{\'e} superlattice with a periodicity that far exceeds the inter-atomic spacing of the constituent crystals. Itinerant electrons in a Fermi sea can be spatially localised by the moir{\'e} potential, leading to the formation of flat bands. The suppressed kinetic energy of the charge carriers relative to their on-site Coulomb repulsion energy, $U$, has led to theoretical predictions \cite{wu2018hubbard,naik2018ultraflatbands,pan2020quantum,padhi2021generalized,morales2021metal,zang2021hartree,zhang2020moire} as well as experimental optical  \cite{tang2020simulation,regan2020mott,shimazaki2020strongly,zhou2021bilayer,liu2021excitonic,xu2020correlated,miao2021strong} and transport \cite{wang2020correlated,li2021continuous,ghiotto2021quantum,li2021charge,huang2021correlated} investigations of strongly correlated electron and hole phases for different TMD homo- and hetero-bilayer systems. In the simplest scenario, the highest flat valence band in a TMD moir{\'e} system can be mapped onto the 2D triangular Hubbard model \cite{wu2018hubbard,morales2021metal,pan2020quantum}. So far, evidence of Hubbard model physics has only been observed experimentally for angle-aligned WSe$_2$/WS$_2$ hetero-bilayers which form a moir{\'e} superlattice due to lattice mismatch \cite{tang2020simulation}. However, the formation and behaviour of exciton-polarons as a function of fractional filling of the moir{\'e} lattice in Hubbard model investigations has yet to be probed.

While strongly correlated phenomena have yet to be observed in moir{\'e} hetero-bilayers formed from WSe$_2$ and MoSe$_2$, the system remains compelling: it is predicted to form flat conduction and valence minibands \cite{ruiz2019interlayer,vitale2021flat} and is an excellent candidate for the realisation of a wide range of strongly correlated states including Wigner crystals \cite{padhi2021generalized,pan2020quantum}, Mott insulators \cite{wu2018hubbard,morales2021metal,naik2018ultraflatbands} and charge-transfer insulators \cite{zhang2020moire}. Compared to TMD hetero-bilayers with different chalcogen atoms, the energetic interplay between Coulomb repulsion and kinetic energy in the WSe$_2$/MoSe$_2$ system is more tunable with relative twist angle due to the small ($0.2 \%$) lattice mismatch \cite{brixner1962preparation}. In addition, the moir{\'e} potential in WSe$_2$/MoSe$_2$ hetero-bilayers has led to the observation of trapping of interlayer excitons at specific atomic registries in the moir{\'e} lattice \cite{yu2017moire,seyler2019signatures,brotons2020spin,baek2020highly,liu2021signatures,brotons2021moir,baek2021optical,wang2021moire}. To date, hetero-bilayer WSe$_2$/MoSe$_2$ remains the only moir{\'e} system to conclusively exhibit exciton trapping. 

Here we optically investigate the formation and behaviour of exciton-polarons, including their charge screening and Coulomb and magnetic interactions, which are formed by intralayer excitons dressed by a Fermi sea localised by a moir{\'e} superlattice in a MoSe$_2$/WSe$_2$ hetero-bilayer. As the Fermi level is tuned, we observe ordering and re-ordering of itinerant carriers into a multitude of correlated states as evidenced by abrupt changes in the oscillator strength, energy, and linewidth of the exciton-polarons. We observe these correlated states at positive (electron) and negative (hole) fractional fillings ($\nu$) of the moiré lattice, including: $\nu = \pm1/3, \pm1/2, \pm2/3, \pm1$, $-5/4$, where $|\nu| = 1$ represents a single carrier per moiré site. We assign the $\pm1$ state to be a Mott \cite{tang2020simulation,regan2020mott,xu2020correlated} or charge-transfer \cite{zhang2020moire} insulator state and the rest to be generalised Wigner crystals \cite{regan2020mott,xu2020correlated,huang2021correlated,liu2021excitonic}. At $\nu = -1$ we observe that the repulsive WSe$_2$ exciton-polaron gains oscillator strength from the attractive, due to the reduced screening of the exciton by carriers in the insulating state. After $\nu = -1$, the oscillator strength is then abruptly transferred to the attractive polaron. Further, we observe filling factor dependent $g$-factors of these positively-charged attractive and repulsive polarons, with a maximum at $\sim$ 1 hole per site. Under the assumption that the $g$-factors are proportional to the magnetic susceptibility of the correlated phases induced by exchange interactions, temperature dependent measurements for $\nu \approx$ $-0.7$ to $-1.3$ reveal an antiferromagnetic spin coupling of the moir{\'e} pinned Fermi-hole sea. The experimental magnetic behaviour is theoretically explored using a model that solves the Heisenberg Hamiltonian for charged-ordered hole states with antiferromagnetic next-neighbour interactions. Our results highlight the behaviour of excitons dressed by a Fermi sea which is spatially ordered in a series of correlated states. This has importance for optical studies of correlated electronic phenomena in 2D materials and demonstrates the potential of MoSe$_2$/WSe$_2$ heterostructures for future investigations and implementations of highly tunable 2D Fermi-Hubbard or Bose-Hubbard models \cite{kennes2021moire}.


\begin{figure*}
    	\begin{center}
    		\includegraphics[scale=0.5]{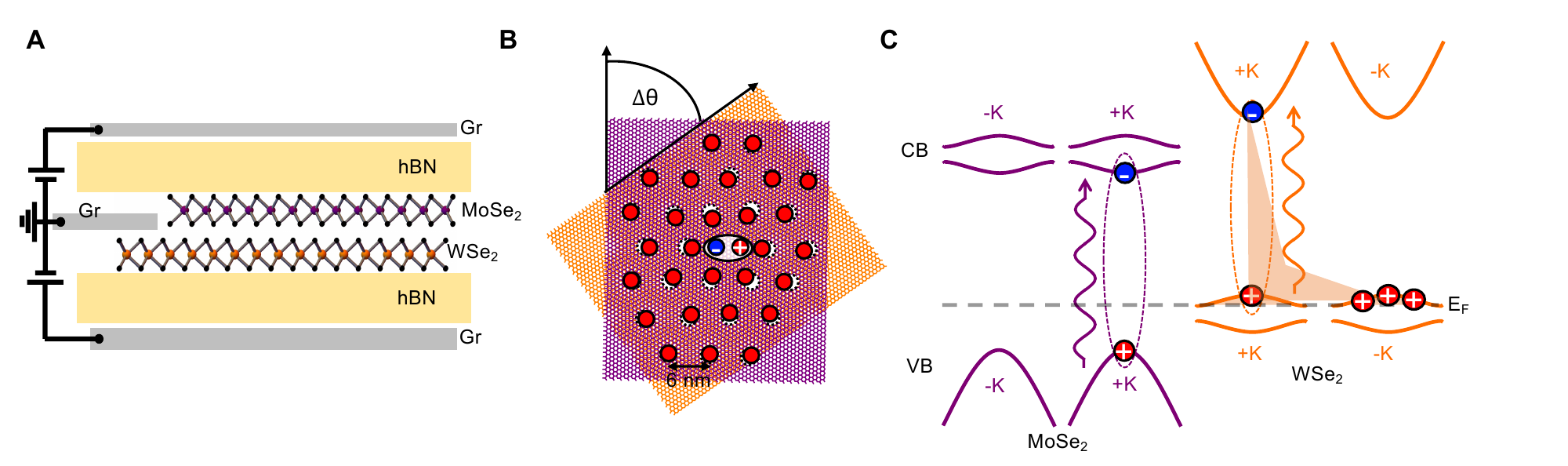}
    	\end{center}
        \caption{\textbf{Device structure and characterisation.} \textbf{(A)} Sketch of the dual-gated WSe$_2$/MoSe$_2$ hetero-bilayer. Graphene layers are used as top, bottom, and hetero-bilayer electrical contacts, while 18-nm-thick hBN layers are used as dielectric spacers \cite{baek2020highly}. \textbf{(B)} Illustration of the top view of a moir{\'e} superlattice with a twist angle $\Delta \theta$ and fractional filling of one hole per moir{\'e} unit cell. Fermi-polarons form between the photo-excited electron-hole pair and charge-ordered Fermi sea. \textbf{(C)} Schematic type-II band structure of the 2$H$-WSe$_2$/MoSe$_2$ hetero-bilayer. Purple and orange curves denote bands from MoSe$_2$ and WSe$_2$, respectively. The vertical wavy arrows represent the photon absorption by intralayer excitons in each monolayer. When the Fermi level ($E_F$) is tuned into the top valence band, the WSe$_2$ attractive polaron forms due to interactions between the exciton in one valley, dressed by holes in the opposite valley.}
        \label{fig1}
\end{figure*}

\section{Results}

\subsection{Device structure and Fermi-polarons}

Figure \ref{fig1}A shows a sketch of our dual-gated hetero-bilayer device, consisting of a monolayer MoSe$_2$ and a monolayer WSe$_2$ vertically stacked with a twist angle ($\Delta \theta$) of $\sim57^{\circ}$. The relative twist angle from perfect 2H stacking (i.e. $\Delta\theta = 60^{\circ}$), estimated from the optical micrograph of the hetero-bilayer (see Materials and Methods) and confirmed by our gate dependent measurements (described later), is beyond the  proposed critical angle for lattice reconstruction \cite{rosenberger2020twist,weston2020atomic}. The hetero-bilayer was encapsulated by hexagonal boron nitride (hBN) layers with nearly identical thicknesses ($\sim18$~nm). Graphene layers act as electrical contacts for the top, bottom and hetero-bilayer gates (see Ref. \cite{baek2020highly} for more details). Moreover, the combination of the layer twist and the lattice mismatch between MoSe$_2$ and WSe$_2$ results in the formation of a triangular moir{\'e} superlattice in our device (see sketch in Fig. \ref{fig1}B) with a period of $\sim6$ nm. This causes a periodic variation in the interlayer hopping that results in a flattening of the conduction and valence bands in the type-II band structure characteristic of TMD hetero-bilayers (see Fig. \ref{fig1}C) \cite{chiu2015determination,wilson2017determination}. The bare intralayer excitons of the constituent TMDs can be probed via absorption spectroscopy. The moir{\'e} lattice carrier concentration is tuned via the application of a gate voltage ($V_g$) between the top/bottom graphene contacts and the hetero-bilayer. As depicted by the schematic in Fig.~\ref{fig1}B, these carriers are spatially ordered in a series of correlated states by the moir{\'e} superlattice, whilst also dressing the photo-excited intralayer exciton to form attractive and repulsive exciton-polaron complexes (as shown in Fig. \ref{fig1}C). 

\begin{figure*}
    	\begin{center}
 		\includegraphics[scale= 1]{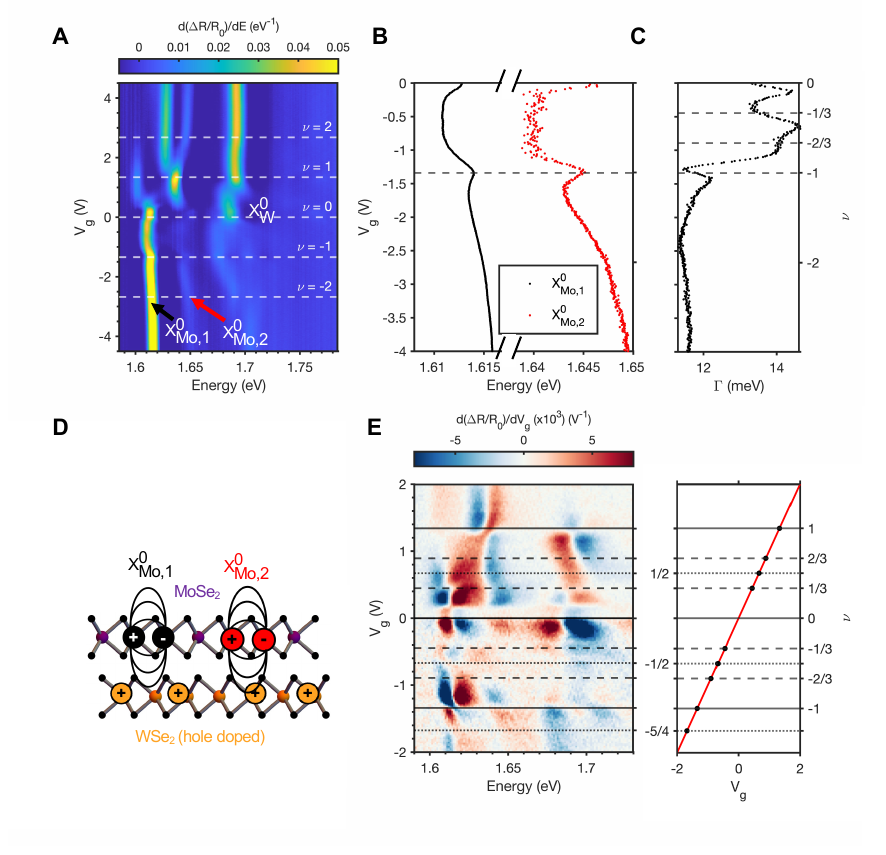}
    	\end{center}
        \caption{\textbf{Correlated electron and hole states in the triangular moir{\'e} superlattice of a 2$H$-WSe$_2$/MoSe$_2$ hetero-bilayer}. \textbf{(A)} Density plot of the first derivative with respect to energy of the differential reflection contrast (d$(\Delta R/R_0)$/d$E$) of the MoSe$_2$ and WSe$_2$ intralayer exciton-polarons in the WSe$_2$/MoSe$_2$ hetero-bilayer. \textbf{(B)} Plot of $V_g$ dependence of the estimated peak positions of the MoSe$_2$ excitons. \textbf{(C)} Plot of the $V_g$ dependence of the estimated linewidth $\Gamma$ of X$^0_{Mo,1}$.  \textbf{(D)} Sketch of the concept of dielectric sensing of correlated hole states. \textbf{(E)} Left: Density plot of the first derivative with respect to gate voltage of the $\Delta R/R_0$ in panel A. d$(\Delta R/R_0)$/d$V_g$ is multiplied by a factor of 2 for $V_g < 0$ for better visualisation. Right: Plot of filling factor $\nu$ against $V_g$.}
        \label{fig2}
\end{figure*}

\subsection{Correlated electronic states}

To investigate doping-dependent phenomena, we perform differential reflection contrast ($\Delta R/R_0$) spectroscopy as a function of $V_g$, where $\Delta R=R_s-R_0$, and $R_s$ ($R_0$) is the intensity of the light reflected by the hetero-bilayer (substrate). Figure~\ref{fig2}A shows the $V_g$ dependence of the first derivative of the reflectance spectra with respect to photon energy (d$(\Delta R/R_0)$/d$E$). The doping dependence of the intralayer exciton-polarons in the MoSe$_2$/WSe$_2$ hetero-bilayer is markedly different to that observed for individual monolayers (see Fig.~S1). At charge neutrality, we observe three excitonic resonances: X$^{0}_{W}$ at higher energy, and two resonances separated by 36 meV in the spectral range corresponding to X$^{0}_{Mo}$, which we label as X$^{0}_{Mo,1}$ (low energy) and X$^{0}_{Mo,2}$ (high energy). We assign the two MoSe$_2$ peaks to be a consequence of the formation of moir{\'e} mini-bands, arising from the band folding at the edges of the reduced Brillouin zone \cite{wu2017topological}. Further, with increasing electron (hole) doping X$^{0}_{W}$ (X$^{0}_{Mo,1}$) dominate the spectrum, as expected for a type-II band alignment \cite{wilson2017determination,hong2014ultrafast}. 


We employ the parallel plate capacitance model to estimate the dependence of the nominal carrier concentration $n$ on the applied $V_g$. Using the density of moir{\'e} sites $n_0$ corresponding to $\Delta \theta \sim 57^\circ$, we estimate the $V_g$-dependent nominal fractional filling $\nu=n/n_0$ of the moir{\'e} lattice (see Materials and Methods). The excitonic features shown in Fig. \ref{fig2}A exhibit strong modulations in their transition energy, linewidth and oscillator strengths for applied voltages close to the nominal $V_g$ values corresponding to $\nu=0$ and $\pm1$. Similar modulations of the excitonic transitions, observed in WSe$_2$/WS$_2$ hetero-bilayers \cite{tang2020simulation,xu2020correlated,jin2021stripe,liu2021excitonic}, have been attributed to the suppressed charge screening originating from the formation of correlated insulator phases at different fractional fillings of the moir{\'e} superlattice. In order to corroborate the presence of a robust moir{\'e} lattice at the same spatial position in our sample where we observe strongly correlated states, we measure the low-temperature (4 K) photoluminescence (PL) spectrum at charge neutrality using confocal spectroscopy, revealing a series of discrete peaks with narrow line-widths ($<$100$~\mu$eV) that demonstrate the existence of an underlying moir{\'e} lattice responsible for the interlayer exciton trapping \cite{seyler2019signatures, brotons2020spin, baek2020highly} (see Fig.~S2). 

Moreover, Fig.~\ref{fig2}A also reveals that each monolayer in the WSe$_2$/MoSe$_2$ hetero-bilayer is capable of sensing the doping-induced changes in their dielectric environment originating from the fractional filling of the other layer, similar to the effects observed using a WSe$_2$ sensor layer in  proximity to a WSe$_2$/WS$_2$ heterostructure \cite{xu2020correlated}. Figure~\ref{fig2}B shows an example of the sensing capabilities of the MoSe$_2$ layer for hole doping of the WSe$_2$ layer: the transition energies of X$^{0}_{Mo,1}$ and X$^{0}_{Mo,2}$ blue-shift and peak at $\Delta V_g=-1.34$~V, consistent with a decrease in the permittivity of the heterostructure arising from the formation of a correlated insulating state at 1 hole per moir{\'e} site in the WSe$_2$ layer \cite{tang2020simulation,liu2021excitonic}. In addition to the modulation in the transition energy, the linewidth of X$^{0}_{Mo,1}$ also presents a clear minimum at $\nu \approx -1$ (see Fig. \ref{fig2}C), which can be understood as the result of reduced charge disorder originating from a correlated insulating state \cite{zhou2021bilayer}. These results demonstrate the potential of intralayer excitons as sensors that can probe the formation of correlated states in the adjacent layer (see sketch in Fig. \ref{fig2}D) and corroborate the calibration of $\nu=\pm1$ in our device. To estimate the $V_g$ values corresponding to other fractional fillings of the moiré lattice, we assume a linear dependence of $\nu$ with $V_g$ and extrapolate from the experimental $V_g$ values determined for one hole/electron per site, as shown in the right panel of Fig. \ref{fig2}E. To increase the sensitivity to doping-induced modulations of the reflectance signal, we plot the first derivative of $\Delta R/R_0$ with respect to $V_g$ (d$(\Delta R/R_0$)/d$V_g$) as a function of $V_g$ (see left panel of Fig. \ref{fig2}E). The d$(\Delta R/R_0)$/d$V_g$ spectrum highlights a series of abrupt changes in the reflected signal at  $\nu=0, \pm1/3, \pm1/2, \pm2/3, \pm1, -5/4,$ (as indicated by the horizontal lines in Fig. \ref{fig2}E), suggesting the formation of correlated states at these fractional fillings of the triangular lattice. These results reveal symmetric loading of carriers, with an identical $\Delta V_g = \pm1.34$~V required to fill the moir{\'e} superlattice with either one electron ($V_g=1.34$ V) or one hole ($V_g=-1.34$ V) per site, respectively. We assign the stable phases at $\nu = \pm1$ to be either Mott \cite{tang2020simulation,regan2020mott,xu2020correlated} or charge-transfer  \cite{zhang2020moire} insulator states and the remaining states to be generalised Wigner crystals \cite{regan2020mott,xu2020correlated,huang2021correlated,liu2021excitonic,padhi2021generalized}.

To gain deeper insight into the strength of the electronic correlations in our system, we investigate the melting temperature of the different correlated states. Figure S9 shows the dependence of d$(\Delta R/R_0)$/d$V_g$ on $V_g$ for temperatures ranging from 4 K to 90 K. With increasing temperature, the abrupt changes in the d$(\Delta R/R_0)$/d$V_g$ spectrum (indicative of correlated state formation in both the electron and hole doping regimes, see Fig. \ref{fig2}E) progressively smooth out until they can no longer be observed at 90 K. We quantitatively estimate a melting temperature of $\sim$55~K for the correlated state at one hole per moir{\'e} site (see Fig.~S10).

\subsection{Exciton-polaron behaviour at one hole per site}

\begin{figure*}
    	\begin{center}	\includegraphics[scale=0.65]{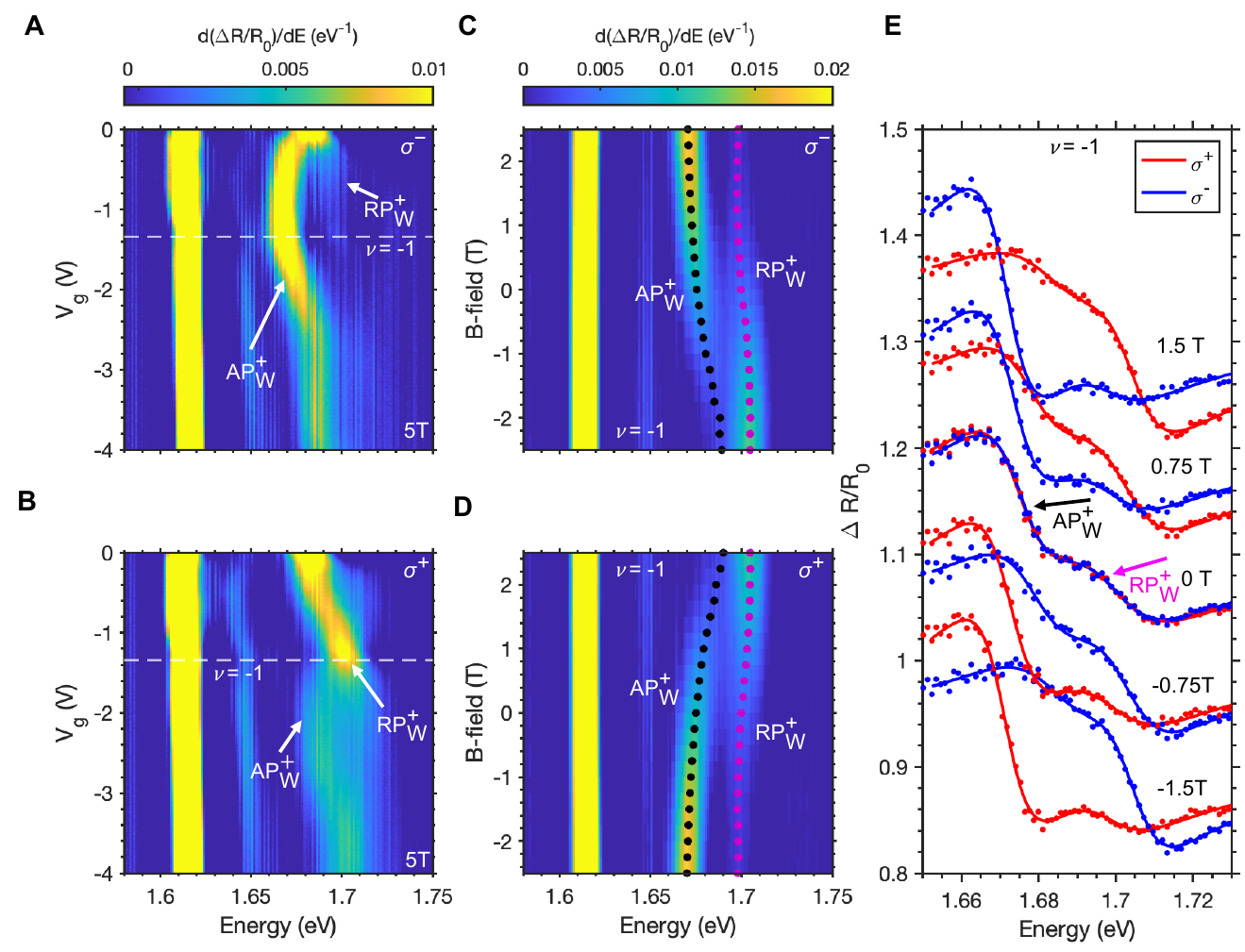}
    	\end{center}
        \caption{\textbf{Exciton-polaron behaviour versus fractional filling of the moir{\'e} lattice} \textbf{(A, B)} $\sigma^-$- (A) and $\sigma^+$-helicity-resolved (B) evolution of the first derivative of $\Delta R/R_0$ with respect to energy (d$(\Delta R/R_0)$/d$E$), as a function of the applied negative $V_g$ (hole doping) under an applied magnetic field of 5 T. \textbf{(C, D)} $\sigma^-$- (C) and $\sigma^+$-helicity-resolved (D) evolution of d$(\Delta R/R_0)$/d$E$ for applied magnetic fields between $-2.5$ T and 2.5 T at $\nu=-1$. The purple and black dots represent the magnetic-field-dependent estimated energies of the RP$_W^+$ and AP$_W^+$, respectively, as extracted from the fits shown in E. \textbf{(E)} Spectra of the bare $\Delta R/R_0$ at different applied magnetic fields (dots) for $\sigma^+$- (red) and $\sigma^-$-polarised (blue) collection. The solid lines represent fits of the experimental $\Delta R/R_0$ to the analytical model described in Materials and Methods.}
        \label{fig3}
\end{figure*}

We now investigate the behaviour of the WSe$_2$ exciton-polarons as the hole fractional filling of the moir{\'e} superlattice is tuned. Figures \ref{fig3}A and \ref{fig3}B show the $\sigma^-$- and $\sigma^+$-helicity-resolved evolution of the d$(\Delta R/R_0)$/d$E$ spectrum, respectively, for negative $V_g$ under an applied magnetic field $B$ of 5 T in Faraday configuration. The intensity colour scale in these figures is saturated to improve the visibility of the WSe$_2$ intralayer repulsive and attractive exciton-polarons (labeled RP$_W^+$ and AP$_W^+$, respectively), while the application of a magnetic field breaks the energy degeneracy between the exciton transitions at $\pm$K, helping to disentangle the behaviour of each excitonic species. As for the monolayer case, at $V_g = 0$ V only the neutral exciton resonance which becomes RP$_W^+$ is present. As the hole fractional filling increases, an additional resonance gains oscillator strength at $\sim$10 meV lower energy, in agreement with the formation of AP$_W^+$ \cite{wang2017probing,van2019probing,courtade2017charged}. 

The helicity-resolved results in Figs. \ref{fig3}A and \ref{fig3}B reveal additional features of the RP$_W^+$ and AP$_W^+$ complexes for fractional hole filling. First, hole doping results in a larger blue-shift of RP$_W^+$ compared to AP$_W^+$, as also observed for excitons interacting with a 2D fermionic sea in ML TMDs \cite{sidler2017fermi,back2017giant,roch2019spin}. Second, the oscillator strength of the RP$_W^+$ resonance shows a non-monotonic behaviour: for small gate voltages it decreases with increasing hole doping, which can be understood as a progressive transfer of oscillator strength from the neutral exciton to the positive trion-like state as the Fermi energy moves deeper into the valence band. However, the RP$_W^+$ resonance regains oscillator strength for hole doping levels corresponding to $\nu \approx -1$. This can be attributed to the suppressed charge screening in the correlated insulating phase forming in the moir{\'e} lattice. For further hole doping ($\nu<-1$), RP$_W^+$ abruptly quenches as the oscillator strength transfers to AP$_W^+$. Third, upon hole doping the $\sigma^+$-polarised transitions of both RP$_W^+$ and AP$_W^+$ appear at a higher energy than their respective $\sigma^-$-polarised transitions, indicative of a positive Zeeman splitting $\Delta E$ (according to the convention based on $\Delta E=E^{\sigma^+}-E^{\sigma^-}$, with $E^{\sigma^{\pm}}$ the energy of the transition with $\sigma^{\pm}$ polarisation). This behaviour contrasts with the negative $g$-factor of exciton-polarons in ML WSe$_2$ based on their spin and valley configurations \cite{forste2020exciton}.  

To gain insight into the origin of the positive Zeeman splitting, we investigate the energies of RP$_W^+$ and AP$_W^+$ as a function of the applied $B$ field for $\nu=-1$. Figures \ref{fig3}C and \ref{fig3}D show the $\sigma^-$- and $\sigma^+$-helicity-resolved evolution of the d$(\Delta R/R_0)$/d$E$ spectrum for applied $B$ fields between $-2.5$ T and 2.5 T at $\nu=-1$. Figure \ref{fig3}E shows linecuts of the bare $\Delta R/R_0$ spectrum at different $B$ fields for $\sigma^+$- (red dots) and $\sigma^-$-polarised (blue dots) collection while the solid lines represent fits from which we estimate the energy, linewidth, and oscillator strength of both RP$_W^+$ and AP$_W^+$ as function of the applied magnetic field (see Materials and Methods for a detailed description of the fitting procedure). The $B$-field-dependent estimated energies of RP$_W^+$ and AP$_W^+$ extracted from the fits are overlayed in Figs. \ref{fig3}C and \ref{fig3}D (purple and black dots, respectively). A large positive Zeeman splitting is clearly observed, which suggests an interaction-enhanced magnetic response of the correlated hole state at $\nu=-1$. Finally, we note that Figs. \ref{fig3}A-\ref{fig3}E reveal a large spin polarisation of RP$_W^+$ and AP$_W^+$ under applied $B$ fields. Such spin polarisation originates from the different effective hole doping in the $\pm$K valleys induced by the large Zeeman splitting. 
\linebreak

\begin{figure*}
    	\begin{center}
    		\includegraphics[scale=0.75]{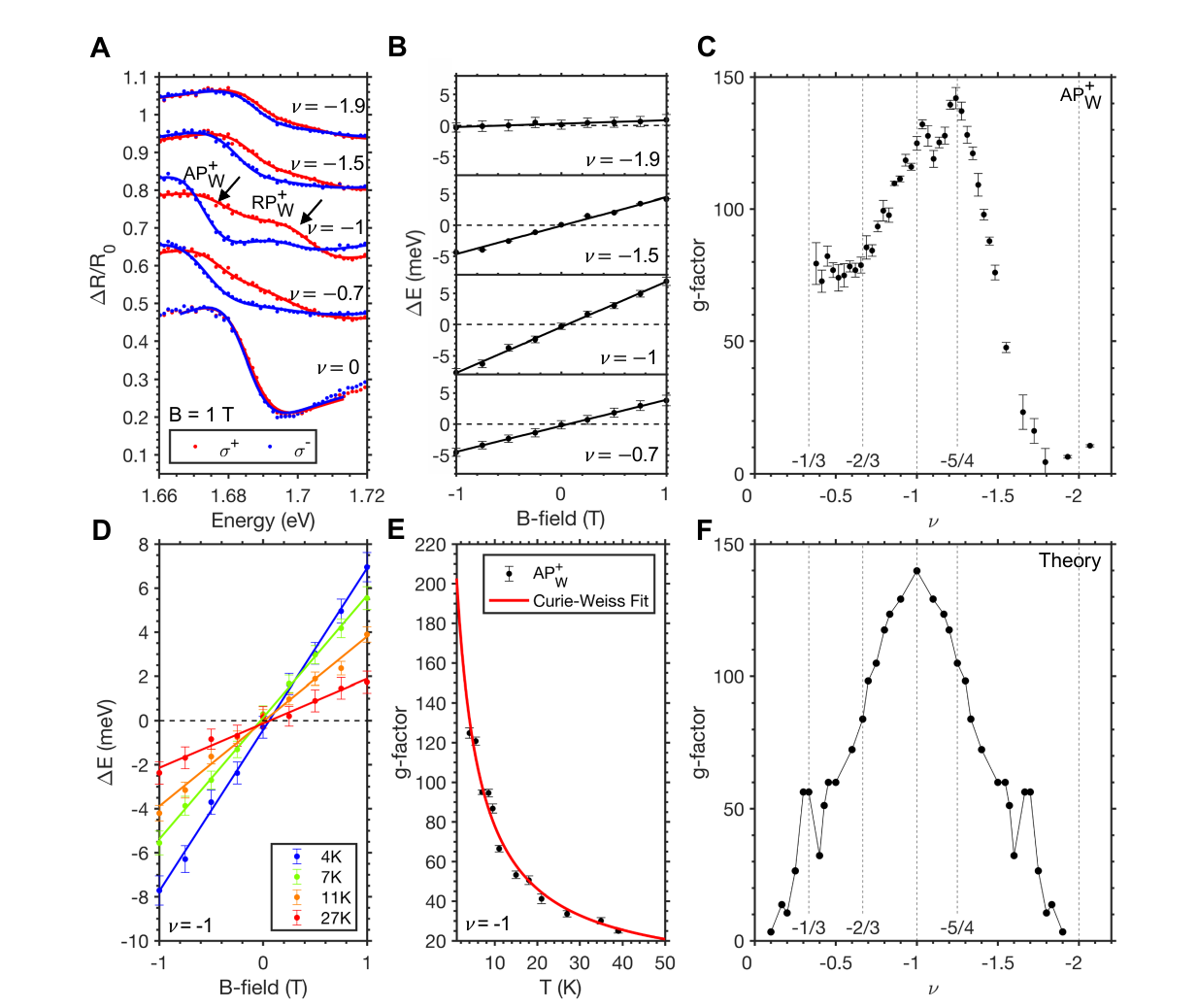}
    	\end{center}
        \caption{\textbf{Exciton-polaron magnetic interactions at different hole fractional fillings} \textbf{(A)} $\sigma^-$- (blue) and $\sigma^+$-resolved (red) $\Delta R/R_0$ spectra at representative hole $\nu$ values under a $B=1$ T. \textbf{(B)} $B$-field-dependent Zeeman splitting of AP$_W^+$ from $-1$ to 1 T at representative hole $\nu$ values. \textbf{(C)} Evolution of the $g$-factor of AP$_W^+$ as a function of the hole $\nu$ extracted from linear fits of $\Delta E$ in the range $|B|\leq1$ T (solid lines in Fig. \ref{fig4}B). \textbf{(D)} Valley Zeeman splitting of the AP$_W^+$ resonance at $\nu=-1$ for different temperatures. \textbf{(E)} Evolution of the measured $g$-factor of AP$^+_W$ as a function of temperature for $\nu=-1$ in the temperature range for which the oscillator strength and linewidth of AP$^+_W$ are sufficient to enable a reliable estimate of the Zeeman splitting. The red solid line represents a fit of the experimental data (black dots) to a Curie-Weiss law from which the Weiss constant is estimated to be $\theta=-4.6\pm0.9$~K. The negative sign of the extracted Weiss constant reveals antiferromagnetic ordering of neighbouring hole spins. \textbf{(F)} Theoretical prediction of the $\nu$ dependence of the $g$-factor based on a model that solves the Heisenberg Hamiltonian for charge ordered hole states with antiferromagnetic exchange interactions between nearest neighbour spins.}
        \label{fig4}
\end{figure*}

\subsection{Magnetic interactions probed by exciton-polarons}

Next, we investigate the fractional-filling-dependence of the Zeeman splitting of the RP$_W^+$ and AP$_W^+$ resonances. Figure \ref{fig4}A shows $\sigma^-$- (blue) and $\sigma^+$-resolved (red) $\Delta R/R_0$ spectra at representative hole $\nu$ values at $B=1$ T. The dots represent experimental data while the solid lines are fits of the experimental $\Delta R/R_0$ to the model described in Materials and Methods, from which we estimate the energy of the resonances. The spectra in Fig. \ref{fig4}A reveal a clear positive $\Delta E$ for all hole doping levels, although with a magnitude that depends strongly on $\nu$. Figure \ref{fig4}B shows the $B$-field-dependent Zeeman splitting of AP$_W^+$ from $-1$ to 1 T at representative hole $\nu$ values. The estimated Zeeman splitting exhibits a linear dependence with $B$ at small fields (i.e., $|B|< 1$ T). We note that RP$_W^+$ shows a similar positive linear dependence with $B$ at small fields, although it saturates at larger $B$ (see Fig. S6). The linear evolution of $\Delta E$ at small $B$ can be associated to an effective exciton valley $g$-factor according to $\Delta E(B) = g\mu_0B$, where $\mu_0$ is the Bohr magneton. Figure \ref{fig4}C shows the evolution of the $g$-factor of AP$_W^+$ as a function of the hole $\nu$ extracted from linear fits of $\Delta E$ in the range $|B|\leq1$ T (solid lines in Fig. \ref{fig4}B). As already inferred from the results in Fig. \ref{fig4}B, the $g$-factor of AP$_W^+$ shown in \ref{fig4}C exhibits a strong dependence on $\nu$, peaking around $\nu\approx-1$, where it reaches a maximum value of $g\sim145$. 

Figure \ref{fig4}D shows the Zeeman splitting of AP$^+_W$ measured for $|B|\leq1$~T at $\nu=-1$ for different temperatures, where we observe the slope of the Zeeman splitting (and therefore the $g$-factor) decreases with increasing temperature. Figure \ref{fig4}E shows the evolution of the measured $g$-factor of AP$^+_W$ as a function of temperature for $\nu=-1$ in the temperature range in which the oscillator strength and linewidth of AP$^+_W$ enable a reliable estimate. We observe that the $g$-factor decreases by a factor $\sim$5 when the temperature increases from 4 K to 39 K. We assume that the interaction-induced enhancement of the attractive polaron $g$-factor is proportional to the magnetic susceptibility of the correlated states (e.g. $\chi \propto g$) and observe that the decrease of $g$-factor with increasing temperature follows a Curie-Weiss law $\chi^{-1}\propto T-\theta$ (red solid line in Fig. \ref{fig4}E), with $T$ being the temperature and $\theta$ the Weiss constant. From the fit in Fig. \ref{fig4}E we estimate a Weiss constant of $\theta=-4.6\pm0.9$~K, which suggests an antiferromagnetic behaviour of the interactions between the localised hole moments for $\nu=-1$. Figure~S11 shows the temperature dependence of the AP$^+_W$ $g$-factor for a range of hole filling factors from $\nu=-0.7$ to $\nu=-1.37$. The Weiss constants extracted from the Curie-Weiss fits are negative for all the explored hole filling factors, suggesting an antiferromagnetic phase for all correlated hole states. We note we observe no magnetic hysteresis in the Zeeman splitting at $\nu = -1$ when the magnetic field is swept from negative to positive values followed by a subsequent positive to negative sweep (see Fig.~S7). In contrast to the large attractive polaron $g$-factor enhancement observed under hole doping, we only observe a modest $g$-factor enhancement in the electron doping regime (see Fig.~S8.).
\linebreak

\section{Discussion}

The extraordinary $g$-factors observed under hole doping can be understood by considering the effect of a magnetic field on a localised hole in the triangular moir\'e superlattice. As a result of exchange interactions with other holes in its environment, such a hole experiences an effective magnetic field which is the sum of the externally applied field and the field induced by the other holes which in turn is proportional to the magnetisation $M$ of the localised hole gas (i.e., $\propto\lambda_X M$, with $\lambda_X$ being a coupling constant).

To calculate the induced field for a given fractional filling, we first determine the configuration of localised carriers that minimises the electrostatic repulsion energy using a simulated annealing technique (see Supplementary text for details). To describe the spin response of this arrangement of charges, we consider a Heisenberg Hamiltonian with distance-dependent antiferromagnetic isotropic exchange interactions $J(r)=J_0 \exp(-r/r_0)$ with $J_0$ denoting the magnitude of the exchange coupling at the characteristic length scale $r_0$. For this Hamiltonian, the induced magnetic field and the corresponding $g$-factor enhancement, $g^\ast/g$, of a localised hole are calculated within mean-field theory (see Supplementary text for a detailed description). To obtain the effective $g$-factor of the attractive exciton-polaron which is probed in our experiments, we assume that its $g$-factor enhancement due to exchange interactions with localised holes is the same as that of a single hole, but that the `non-enhanced' $g$-factor $g$ can be different. The experimental value of this non-enhanced $g$ is unknown since the $g$-factors of exciton-polarons are dependent on paramagnetic interactions and phase-space filling effects as carrier concentration is changed, even for monolayer TMDs \cite{back2017giant}. Therefore, we treat $g$ as an adjustable parameter choosing its value such that the calculated and experimentally measured $g$-factors agree at  $\nu = -1$.

Figure~\ref{fig4}F shows that the filling dependence of the calculated $g$-factor is in good qualitative agreement with the experimental results. Specifically, it reaches a maximum at $\nu=-1$, where the average number of occupied moir\'e sites around the localised hole is largest and the strong exchange interactions between neighboring spins give rise to a large effective magnetic field. The model also captures the plateau-like feature between $\nu=-1/3$ and $-2/3$. In contrast to the experimental findings, however, the calculated $g$-factor is symmetric around $\nu=-1$. Further theoretical work is required to understand this discrepancy. 

Finally, we estimate $U/t$ in our device. The antiferromagnetic coupling between neighbouring spins due to the kinetic exchange mechanism can be estimated as $J\approx-t^2/U$, where $t$ is the hopping amplitude between neighbouring moir{\'e} lattice sites. Using the value of $\theta$ at $\nu = -1$ we estimate $J\approx-0.4$ meV. By combining $J$ with the estimated melting temperature of the correlated state at $\nu=-1$ ($\sim$55~K) we obtain $U/t\approx3.5\pm0.4$. This experimental value agrees well with predicted values for MoSe$_2$/WSe$_2$ heterostructures with stacking angles $\sim3^{\circ}$ and $\sim57^{\circ}$ \cite{wu2018hubbard}.

\section{Conclusion}
Our results illustrate the properties of exciton-polarons in the presence of correlated states in moir{\'e} heterostructures. Using the changes in energy, oscillator strength, and linewidth of intra-layer excitons dressed by itinerant carriers occupying narrow electronic moir{\'e} bands in a MoSe$_2$/WSe$_2$ hetero-bilayer, we observe the formation of correlated electron and hole states at a multitude of fractional fillings of the moir{\'e} lattice. Upon hole doping, the WSe$_2$ attractive polaron transfers oscillator strength back to the repulsive polaron branch at 1 carrier per site, demonstrating the reduced screening of the exciton by the free charges in the presence of the insulating state. In addition, we observe the magnetic interactions within the correlated hole states via both the attractive and repulsive WSe$_2$ exciton-polarons, which exhibit enhanced Zeeman splittings due to exchange interactions with the moir{\'e} pinned carriers. Through temperature dependent measurements, the magnetic ordering of the correlated holes is shown to be antiferromagnetic in the range $\nu = -0.7$ to $\nu = -1.37$, and the $U/t$ ratio of our device is estimated to be $\sim$3.5. Further investigations could exploit the small lattice mismatch between MoSe$_2$/WSe$_2$, which enables a highly tunable moir{\'e} period, to simulate condensed matter phase diagrams over a large range of $U/t$ ratios. Our observation of the formation of flat electronic bands compliments recent reports of moir{\'e} trapped interlayer excitons in MoSe$_2$/WSe$_2$ hetero-bilayers \cite{baek2021optical,baek2020highly,brotons2020spin,liu2021signatures,seyler2019signatures,wang2021moire,brotons2021moir} and highlights the exciting prospects to investigate Fermi-Hubbard and Bose-Hubbard physics in this system.  

\section{Acknowledgements}

We thank Mikhail M. Glazov for fruitful discussions. This work was supported by the EPSRC (grant nos. EP/P029892/1 and EP/L015110/1), the ERC (grant no. 725920) and the EU Horizon 2020 research and innovation program (grant agreement no. 820423). M.B.-G. is supported by a Royal Society University Research Fellowship. B.D.G. is supported by a Wolfson Merit Award from the Royal Society and a Chair in Emerging Technology from the Royal Academy of Engineering. V.V and J.L. acknowledge funding from the EPSRC (grant no. EP/S025324/1). K.W. and T.T. acknowledge support from the Elemental Strategy Initiative
conducted by the MEXT, Japan (Grant Number JPMXP0112101001) and JSPS KAKENHI (Grant Numbers 19H05790, 20H00354 and 21H05233). 

\bibliography{Correlated_states_biblio}

\end{document}